\documentclass[doublecol]{epl2} 
\usepackage{mathtools} 
\usepackage{amsfonts} 

\begin{document}

\title{Stochastic analysis of chemical reactions in multi-component \\ interacting systems at criticality}

\shorttitle{Chemical reactions in multi-component interacting systems} 
\author{Reda Tiani \inst{1} \and Uwe C. T\"auber \inst{1,2}}
\shortauthor{Reda Tiani and Uwe C. T\"auber}
\institute{\inst{1} Department of Physics \& Center for Soft Matter and Biological Physics (MC 0435), \\
  Robeson Hall, 850 West Campus Drive, Virginia Tech, Blacksburg, Virginia 24061, USA
  \footnote{\email email: tiani@vt.edu, tauber@vt.edu} \\
  \inst{2} Faculty of Health Sciences, Virginia Tech, Blacksburg, Virginia 24061, USA}

\abstract{
We numerically and analytically investigate the behavior of a non-equilibrium phase transition in the second Schl\"ogl 
autocatalytic reaction scheme. 
Our model incorporates both an interaction-induced phase separation and a bifurcation in the reaction kinetics, with 
these critical lines coalescing at a bicritical point in the macroscopic limit. 
We construct a stochastic master equation for the reaction processes to account for the presence of mutual particle 
interactions in a thermodynamically consistent manner by imposing a generalized detailed balance condition, which 
leads to exponential corrections for the transition rates. 
In a non-spatially extended (zero-dimensional) setting, we treat the interactions in a mean-field approximation, and
introduce a minimal model that encodes the physical behavior of the bicritical point and permits the exact evaluation of 
the anomalous scaling for the particle number fluctuations in the thermodynamic limit. 
We obtain that the system size scaling exponent for the particle number variance changes from $\beta_0 = 3/2$ at the 
standard non-interacting bifurcation to $\beta = 12/7$ at the interacting bicritical point. 
The methodology developed here provides a generic route for the quantitative analysis of fluctuation effects in chemical
reactions occurring in multi-component interacting systems.}

\maketitle

\section{Introduction} 

Most studies addressing the stochastic analysis of chemically reacting systems consider configurations of point-like ``ideal''
particles devoid of mutual forces \cite{Van_Kampen_1976, Nicolis_1977, 
Nicolis_book_1977, Nicolis_1979, Mansour_1981, Van_Kampen_book_1981, Kuzovkov_1988,
Ovchinnikov_book_1989, Hinrichsen_2001, Odor_2004, Tauber_2005, Henkel_book_2008, 
Krapivsky_book_2010, Tauber_book_2014, Lindenberg_book_2019, Gaspard_book_2022}.
The mathematical characterization of such idealized systems, defined by the absence of any interaction Hamiltonian, was 
essential to forge our understanding of fluctuation and correlation effects in systems where chemical reactions are involved, 
and hence for the progress of non-equilibrium statistical mechanics \cite{Kuzovkov_1988, 
Ovchinnikov_book_1989, Hinrichsen_2001, Odor_2004, Tauber_2005, Henkel_book_2008, 
Krapivsky_book_2010, Tauber_book_2014, Lindenberg_book_2019, Gaspard_book_2022}. 
Of particular interest has been the investigation of nonlinear dynamical systems since they generically exhibit similar critical
phenomena at their steady state bifurcation points as are observed at continuous phase transitions in thermal equilibrium 
\cite{Hinrichsen_2001, Odor_2004, Tauber_2005, Henkel_book_2008, Tauber_book_2014}. 
A central result concerning fluctuation effects in such systems is the breakdown of the central limit theorem at bifurcation
points, as, e.g., demonstrated in the second Schl\"ogl model \cite{Schlogl_1972} for which the particle number variance at 
the bistability point scales with system size $\langle (\delta n)^2 \rangle \sim \Omega^{\beta_0}$ with a ``non-classical'' 
exponent $\beta_0 = 3/2$ in the thermodynamic limit \cite{Van_Kampen_1976, Nicolis_1977, Nicolis_1979}.

In this letter, we revisit the stochastic approach of reactive systems when the assumption of \textit{ideality} is relaxed, 
driven by both practical and fundamental considerations. 
From an applied perspective, our work is motivated by the observation that ideality is rather an exception than a rule. 
In living cells, proteins are large and operate in crowded molecular environments, thus experiencing excluded-volume 
interactions, while the (weak) van-der-Waals interactions with the surrounding solvent molecules are vital for the 
stabilization of protein structure and therefore the maintenance of biochemical activity \cite{Yu_2016}. 
Biological organisms, along with supramolecular chemical systems \cite{Odriozola_2013} and soft condensed matter 
\cite{Mikhailov_book_2017}, constitute the most abundant system classes encountered in Nature, and are all intrinsically 
strongly \textit{non-ideal}. 
In particular, considerable effort has been devoted recently to the understanding of chemical reactions of interacting 
particles due to their importance for (liquid-liquid) phase separation occurring inside biological cells 
\cite{Alberti_2017, Weber_2019, LiY_2022, Zwicker_2022}. 
These processes induce the formation of tiny (micron-scale) droplets that aid cells in organizing biochemical reactions in 
space and time \cite{Brangwynne_2015, Shin_2017}. 
Novel scenarios based on this active interplay between the chemistry and the strongly interacting cellular environment  
have recently been envisioned to explain the complexity and origin of life on earth \cite{Zwicker_2022}.

From a theoretical point of view, promising results were previously obtained on the grounds of macroscopic, deterministic 
dynamical laws for reactive processes occurring in interacting systems. 
Deviations from the classical law of mass action as well as from Fickian diffusion in kinetic and transport equations can, for 
example, lead to multiple equilibrium states in \textit{closed} reactive systems \cite{Othmer_1976}, stable limit cycles in 
\textit{non-autocatalytic} polymer reactions \cite{Li_1982}, the enhancement of bifurcation and instability scenarios 
\cite{Li_1981_a, Li_1981_b, Aslyamov_2023}, and the appearance of novel critical phenomena associated with the 
emergence of multi-critical points in phase-separating systems \cite{Li_1981_a}. 
Yet all these findings were based on pure mean-field analyses. 
This calls for a generalization to properly account for the influence of (internal) fluctuations and both reaction- and
  interaction-induced correlations that often play a crucial role in the vicinity of critical points, and therefore in the 
spontaneous generation of dissipative spatio-temporal structures \cite{Nicolis_book_1977}. 

In this context, the aim of this letter is to highlight the effect of fluctuations in multi-component interacting systems subject
to chemical reactions. 
To this end, we revisit the particle number variance and its scaling with system size in an interacting, zero-dimensional
version of the (autocatalytic) second Schlögl model \cite{Schlogl_1972}, where we fully account for  the intrinsic reaction 
stochasticity, but treat the mutual particle interactions in a coarse-grained mean-field approximation.
In the following, we first define our model system and review its mean-field solutions in the presence of averaged 
interactions.
We then introduce a minimal model that faithfully encapsulates the qualitative critical behavior of the original stochastic
reacting system, yet maintains the mean-field interaction character. 
Next we investigate the effect of fluctuations as encoded in the master equation representing stochastic Markovian 
birth-death processes, which we utilize to both analytically and numerically evaluate the particle number variance in the 
thermodynamic limit at criticality and at the steady state. 
In our final section, we draw relevant conclusions and comment on promising prospects for future investigations.

\section{Interacting reacting particle model and mean-field limit}

\begin{figure}[]						
\begin{center}
\includegraphics[width=5.8cm,height=5.8cm,keepaspectratio]{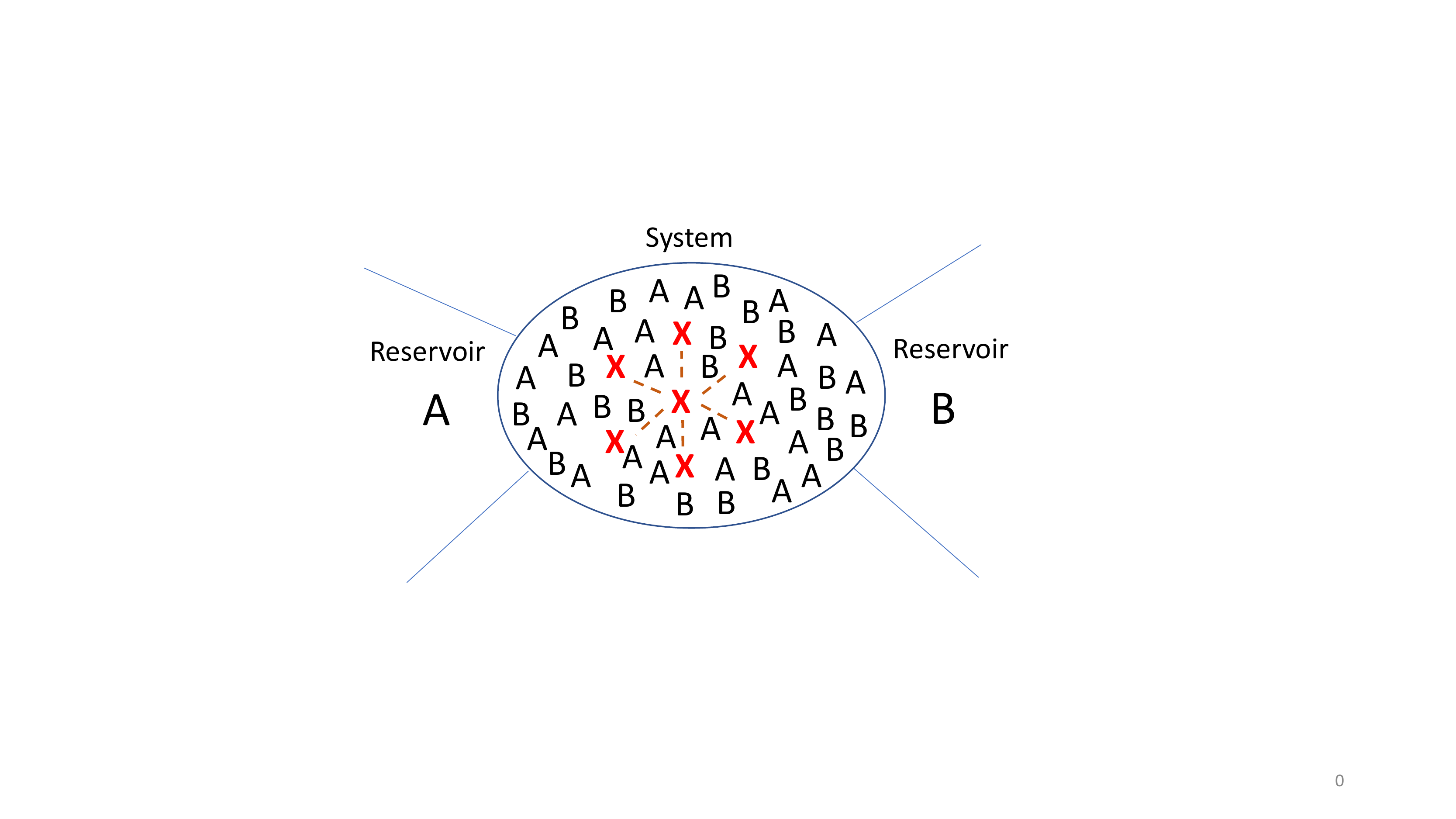}
\caption{Sketch of the interacting and reacting particle system: 
	Each particle of type X feels an average mean-field interaction potential (represented as dotted lines) with other 
	particles of the same species.
	The particles chemically react according to the second Schl\"ogl model scheme (\ref{eq:1}); the reservoirs for species 
      A and B serve to maintain their densities $c_\textrm{A}$ and $c_\textrm{B}$ constant.} 
\label{fig:fig_1}
\end{center}
\end{figure}

Our model system is schematically depicted in fig.~\ref{fig:fig_1}. 
We consider a system composed of three different chemical species, A, B, and X. 
The densities of ``chemostated'' species A and B are kept fixed and controlled by their respective particle reservoirs, 
while we are interested in the temporal evolution of the intermediary species $X$ in this second Schl\"ogl model 
\cite{Schlogl_1972},
\begin{equation}
	\mathrm{A + 2X} \xrightleftharpoons[k_{-1}]{k_{+1}} \mathrm{3X} \ , \quad  
	\mathrm{X} \xrightleftharpoons[k_{-2}]{k_{+2}} \mathrm{B}\ ,
 \label{eq:1}
\end{equation}
where $\{k_{+1}, k_{-1}, k_{+2}, k_{-2}\}$ denotes the set of kinetic reaction rates serving as model parameters. 
In ideal (highly dilute) systems, the particles experience no mutual interactions. 
Here, we relax this assumption by subjecting each particle to a coarse-grained average (mean-field) potential generated by 
the other reactants, as illustrated in fig.~\ref{fig:fig_1} for a particle of type X. 

As shown in the Supplementary material (SM), appendix~A,\footnote{See the SM for the derivation of the mean-field 
(\ref{eq:16}) and stochastic (\ref{eq:32}) models and of the analytical expression for the particle number variance 
(\ref{eq:36}). 
The SM includes refs.~\cite{Groot_book_1984, Hanggi_1990, Li_1981_a, Weber_2019, Zwicker_2022, 
Schwabl_book_2006, Kardar_book_2007, Schmittmann_1990, Schmittmann_book_1995, Julicher_1997, 
Marro_book_1999, Seifert_2005, Pietzonka_2014, Miangolarra_2023, Nicolis_1977}.} 
under the assumption of local thermodynamic equilibrium, the (near-equilibrium) mean-field rate equation governing the 
number density $c(t)$ of particles of species X at time $t$ becomes 
\cite{Hanggi_1990, Groot_book_1984, Schwabl_book_2006, Kardar_book_2007},
\begin{equation}
\begin{split}
	\frac{\partial c}{\partial t} &= D \nabla \left[ (1+wc) \nabla c \right] + k_{+1} \, c_\textrm{A} \, [c \, \gamma(c)]^2 
	\\ &\quad - k_{-1} \, [c \, \gamma(c)]^3 - k_{+2} \, c \, \gamma(c) + k_{-2} \, c_\textrm{B} \ ,
\label{eq:16}
\end{split}
\end{equation}
where $w$ represents the mean-field potential (of unit volume) which plays the role of the interaction parameter of the 
model, $\gamma(c) = \mathrm{exp}(wc)$ is the (composition-dependent) functional correction due to the mean-field 
interactions, and $n_\textrm{A}$, $n_\textrm{B}$, and $D$ denote densities of species A, B and the diffusion coefficient, 
all assumed to be constant). 
We shall take the particles of type X to be mutually attractive, $w < 0$: 
This will turn out to be a necessary condition for the existence of an unmixing transition or phase separation, and hence for 
the emergence of a bicritical point.

Equation~(\ref{eq:16}) is built on the flux--force relations of non-equilibrium thermodynamics and on the transition state
theory of reaction-limited processes consistent with a generalized detailed balance condition to evaluate the diffusion and 
reaction terms, respectively \cite{Hanggi_1990, Groot_book_1984}. 
The mean-field interactions have been incorporated through the partition function of an interacting gas system. 
Thus, eq.~(\ref{eq:16}) represents a basic physical model for the critical behavior of interest here. 
It closely resembles the setup in the pioneering work of Li and Nicolis \cite{Li_1981_a}, who considered a regular solution 
with small interaction parameter $w$, whereas our present model does not rely on this assumption.

Note that the density-dependent effective diffusion coefficient, $D(c) = D \, (1+wc)$ in eq.~(\ref{eq:16}), vanishes when 
$c = - 1 / w$. 
If $c \le -1/w$, $D(c) \ge 0$, and diffusion plays no destabilizing role. 
However, $D(c) < 0$ for $c > -1 / w$, whence a small perturbation in the density field may grow.
A diffusive instability arises and additional scalar terms that include higher gradients of the form $\left( \nabla c \right)^2$ 
and $\nabla^2 c$ are required in the density expansion of the chemical potentials to stabilize the system, \textit{c.f.}
eqs.~(11)--(13) in the SM .
These terms in turn describe the formation of spatial domains as the system is driven inside the spinodal region 
\cite{Cahn_1958, Li_1981_a}. 
Consequently, $c_{cr} = -1 / w > 0$ corresponds to the value of the density at the unmixing point of phase separation, 
which can indeed only occur for attractive forces. 

Specifically, we seek homogeneous steady-state solutions of eq.~(\ref{eq:16}), for which
\begin{equation}
	k_{+1} \, c_\textrm{A} [c \,\gamma(c)]^2 - k_{-1} [c \, \gamma(c)]^3 - k_{+2} \, c \, \gamma (c) 
	+ k_{-2} \, c_\textrm{B} = 0 \ ,
\label{eq:17}
\end{equation}
\noindent
where $c$ now denotes the steady-state density. 
To solve this nonlinear eq.~(\ref{eq:17}), we follow ref.~\cite{Li_1981_a} to define 
$c \, \gamma(c) = q \left[ 1+z(c) \right]$ with $q \ge 0$ and a real-valued function $z(c)$; and introduce the following 
scaled parameters $k_{+1} \, c_\textrm{A} / k_{-1} = 3 q$, $k_{+2} / k_{-1} = (3 + \lambda) q^2$, and 
$k_{-2} \, c_\textrm{B} / k_{-1} = (1+\lambda') q^3$, with $\lambda \ge -3$, $\lambda' \ge -1$, whereupon 
eq.~(\ref{eq:17}) reduces to 
\begin{equation}
	z(c)^3 + \lambda \, z(c) + \lambda - \lambda' = 0 \ .
\label{eq:18}
\end{equation}
In general, eq.~(\ref{eq:18}) describes a discontinuous transition; the presence of a pitchfork bifurcation, \textit{i.e.}, a 
continuous phase transition requires the inversion symmetry $z(c) \leftrightarrow -z(c)$.
Hence for this special situation necessarily $\lambda = \lambda'$; if $\lambda \ge 0$, $z_0 (c) = 0$ is the only real 
solution of eq.~(\ref{eq:18}), while for $\lambda < 0$, there exist two additional real roots 
$z_{\pm} = \pm \sqrt{-\lambda}$. 

\begin{figure}[]						
\begin{center}
\includegraphics[width=8cm,height=8cm,keepaspectratio]{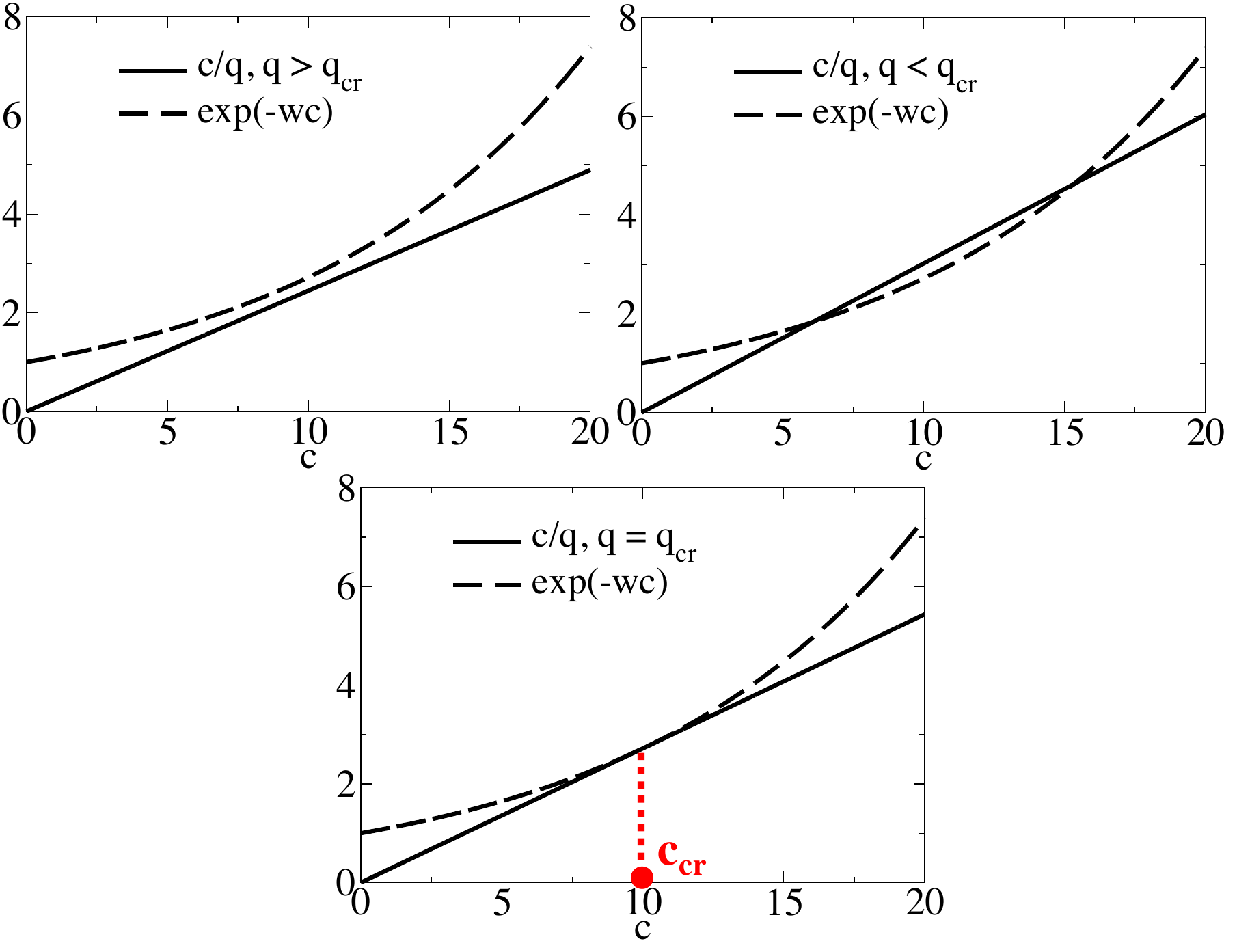}
\put(-138,103){$\displaystyle (a)$}
\put(-24,103){$\displaystyle (b)$}
\put(-76,17){$\displaystyle (c)$}
\caption{Steady-state solutions for $\lambda = 0$ and (a) $q > q_{cr}$, (b) $q < q_{cr}$, and (c) $q = q_{cr}$, with
	$q_{cr} = - 1 / (\mathrm{e} \, w)$ (plotted for $w = - 0.10$). 
	In (c), the two (solid/dashed) functions on both sides of (\ref{eq:19}) intersect at the critical point for the density 
	$c_{cr} = -1 / w$.} 
\label{fig:fig_2}
\end{center}
\end{figure}
We now focus on the multi-critical behavior emerging at $\lambda = \lambda' = 0$.
This triply degenerate case $z_0(c) = 0$ yields $c \, \gamma (c) = q$, \textit{i.e.},
\begin{equation}
	c  / q = \mathrm{exp}(- w c) \ .
\label{eq:19}
\end{equation}
Figures~\ref{fig:fig_2}(a)-(c) illustrate that eq.~(\ref{eq:19}) admits (a) no, for $q>q_{cr}$; (b) two, for $q<q_{cr}$; 
or (c) a single steady-state solution(s) at $q=q_{cr}$, respectively, where $q_{cr}$ is a particular (critical) value of $q$ 
that solely depends on the interaction parameter $w$. 
Indeed, at $q = q_{cr}$, we note that both curves $c / q$ and $\mathrm{exp}(- w c)$ ($w<0$) assume identical 
values and slopes at $c = c_{cr}$, i.e., $c_{cr} / q_{cr} = \mathrm{exp} \left( - w c_{cr} \right)$ and 
$1/q_{cr} = -w \, \mathrm{exp} \left( - w c_{cr} \right)$; consequently,
\begin{equation}
	c_{cr} = - 1 / w \ , \quad q_{cr} = -  1 / (\mathrm{e} \, w) \ .
\label{eq:22}
\end{equation}
We conclude that the reactive critical point is governed by two control parameters $\{\lambda, q\}$ with critical values 
$\{0, - 1 / (\mathrm{e} \, w) \}$, with associated density $c_{cr} = -1 / w$.
Since this is also precisely the density value at the phase separation or unmixing critical point in our model, 
$\{\lambda_{cr}, q_{cr} \}$ constitutes a bicritical point at which the reaction kinetics bifurcation and the phase separation 
transition coincide.

\begin{figure}[]						
\begin{center}
\includegraphics[width=5.8cm,height=5.8cm,keepaspectratio]{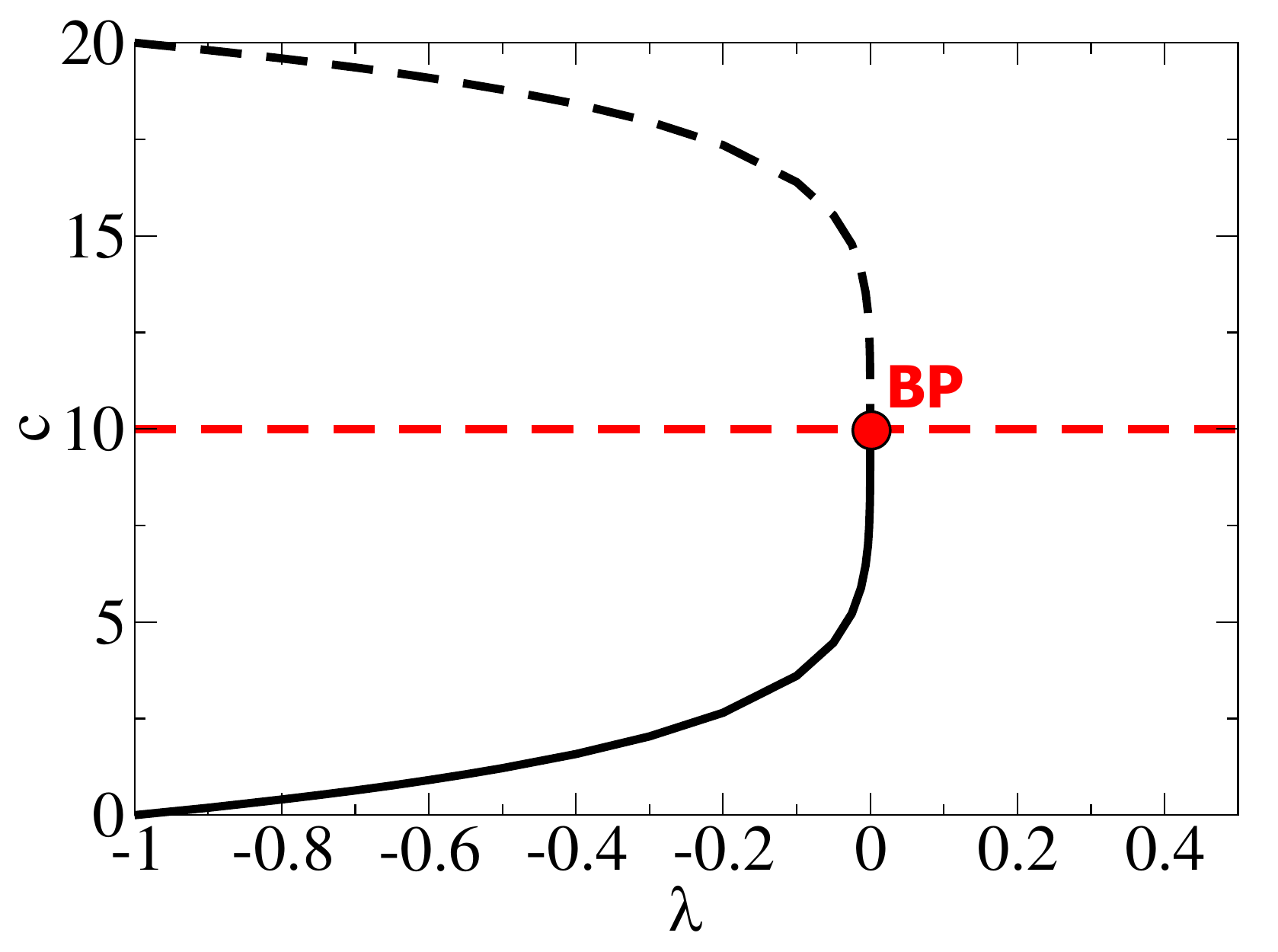}
\caption{Steady-state solutions as a function of $\lambda$ at $q = q_{cr}$, where $q_{cr} = - 1 / (\mathrm{e} \, w)$ 
	(shown for $w = - 0.10$). 
	The solid and dashed lines represent the stable and unstable steady solutions of eq.~(\ref{eq:18}), respectively. 
	For $\lambda = 0$, the solution becomes unique, $c_{cr} = -1 / w$, which defines the model's bicritical point.} 
\label{fig:fig_3}
\end{center}
\end{figure}
The characteristic phase diagram for the stationary density $c$ as a function of $\lambda = \lambda'$ and at $q=q_{cr}$ 
can be obtained by solving eq.~(\ref{eq:18}) numerically, see fig.~\ref{fig:fig_3}. 
For $\lambda = \lambda' < 0$, the steady-state solution $c = -1/w$ bifurcates into a lower ($c < -1/w$) stable and an 
upper ($c > -1/w$) unstable branch, as is readily established by linear stability analysis. 
To obtain further analytical insights close to the bicritical point, we define the order parameter as $\phi = c + 1 / w$. 
From eq.~(\ref{eq:19}) at $q = q_{cr}$ as $\lambda \to 0^-$, we deduce that $\phi(\lambda) \sim \lambda^{1/4}$. 
In contrast, at the bifurcation point as $\lambda \to 0^-$, $\phi(\lambda) \sim \lambda^{1/2}$. 
The critical order parameter exponent thus crosses over from $1/2$ to $1/4$ as the system is driven from the reactive 
bifurcation to the bicritical point.
 
We next investigate the effect of (internal) fluctuations in our (mean-field) interacting system by evaluating the particle 
number variance at the bicritical point. 
Our stochastic approach is based on a master equation generalized to account for the effects of an interaction potential. 
The resulting stochastic dynamical equation is difficult to analyze analytically as the presence of such a potential introduces 
transitions rates with exponential functions that render the use of the traditional methodology such as generating functions
and perturbative approaches more challenging. 
In the following sections, we shall demonstrate that analytic progress is however possible by means of introducing and
solving a minimal model that displays the same qualitative features as the full (exponential) model.

\section{Minimal model and irrelevant terms}

To construct a minimal model that preserves the qualitative information pertaining to the ensuing critical behavior, we 
seek to replace the exponential density dependence in the interaction factor $\gamma (c) = \mathrm{exp}(wc)$ with a 
polynomial.
Hence we proceed with a Taylor expansion to order $p \in \mathbb{N}$ and insert it into eq.~(\ref{eq:19}), which gives 
$q / c = 1+ w c +  (w c)^2 / 2 + \ldots + (w c)^p / p!$.
In order to qualitatively recover the different scenarios of fig.~\ref{fig:fig_2}, the right-hand side must remain a 
monotonically increasing (convex) function of $c$. 
Since $w < 0$, this enforces that the Taylor sum may only terminate at an odd power $p = 2l + 1$:
$q / c = 1+ \sum_{j=1}^{2l+1} (w c)^j/j! +\mathcal{O}(c^{2l+2})$.
The relevant minimal information is already retained setting $l = 0$ or $p = 1$, \textit{i.e.}, by keeping only the two 
lowest-order terms of the expansion.
Additional contributions would merely shift the critical parameter values relative to the simplest scenario, and of course to 
the original exponential density dependence. 
Thus we obtain to order $c$
\begin{equation}
	\frac{c}{q} = \frac{1}{1+ w c} \ ,
\label{eq:24}
\end{equation}
with the density constraint $c < -1 / w$.
As developed for the original model, the left- and right-hand sides of eq.~(\ref{eq:24}) and their derivatives must be
 equal at criticality, yielding 
\begin{equation}
	c_{cr} = - 1 / (2 w) \  , \quad q_{cr} = - 1 / (4 w) \ .
\label{eq:25}
\end{equation}
We conclude that, in this minimal model, the bicritical values of the control parameters $\{\lambda, q\}$ are 
$\{0, - 1 / (4 w) \}$ and the critical density becomes $c_{cr} = - 1 / (2 w)$.

\section{Generalized chemical master equation}

In order to extend our analysis beyond the deterministic mean-field approximation, we describe an elementary reactive 
event as a death-birth Markov process. 
As shown in the SM, appendix B, under well-mixed conditions (no diffusive transport), the probability $P(n,t)$ to find a 
configuration with $n$ particles of type X subject to the reactions (\ref{eq:1}) obeys the continuous-time stochastic
master equation
\begin{equation}
\begin{split}
	&\frac{\partial P(n,t)}{\partial t} = 
		k_{+2} \, (n+1) \, \mathrm{exp}\!\left( \frac{w}{\Omega} n \right) P(n+1,t) \\
	&\ + \frac{k_{+1}}{\Omega^2} \, n_\textrm{A} (n-1) (n-2) \, \mathrm{exp}\!\left[ \frac{w}{\Omega} (2n-5) \right] 
		P(n-1,t) \\
	&\ + \frac{k_{-1}}{\Omega^2} \, (n+1) n (n-1) \, \mathrm{exp}\!\left[ \frac{3w}{\Omega} (n-1) \right] P(n+1,t) \\
	&\ + k_{-2} n_B \, P(n-1,t) - k_{-2} \, n_\textrm{B} \, P(n,t)  \\
	&\ - \frac{k_{+1}}{\Omega^2} \, n_\textrm{A} n(n-1) \, \mathrm{exp}\!\left[ \frac{w}{\Omega} (2n-3) \right] P(n,t) \\
	&\ - \frac{k_{-1}}{\Omega^2} \, n (n-1) (n-2) \, \mathrm{exp}\!\left[ \frac{3w}{\Omega} (n-2) \right] P(n,t) \\
	&\ - k_{+2} \, n \, \mathrm{exp}\left[ \frac{w}{\Omega} (n-1) \right] P(n,t) \ .
\label{eq:32}
\end{split}
\end{equation}
As in the corresponding mean-field description, the form of this master equation relies on transition-state theory and a 
generalized detailed balance condition to fix the forward and backward transition rates associated with the two reversible 
reactive events (\ref{eq:1}); see the SM, appendix B \cite{Hanggi_1990, Schmittmann_1990, Schmittmann_book_1995, 
Julicher_1997, Marro_book_1999, Seifert_2005, Pietzonka_2014, Miangolarra_2023}. 
The derivation of eq.~(\ref{eq:32}) still involves averaging the interaction part of the Hamiltonian over the volume $\Omega$, 
giving rise to the mean-field interaction potential $w$.
Yet the death-birth Markov approach fully encapsulates the stochastic nature of the reactive processes. 
This allows the present model to remain zero-dimensional (no spatial extension), governed by transition rates that are
independent of spatial configurations and locally varying interactions. 
Physically, eq.~(\ref{eq:32}) should be viewed as a first step towards describing the interplay between the stochasticity of
chemical reactions and the effects of molecular interactions that extends beyond the mean-field description (\ref{eq:16}), and 
moreover generalizes stochastic models previously derived in the literature that address only ideal systems ($w = 0$) 
\cite{Van_Kampen_1976, Nicolis_1977, Nicolis_1979, Kuzovkov_1988, Ovchinnikov_book_1989, Hinrichsen_2001, 
Odor_2004, Tauber_2005, Henkel_book_2008, Krapivsky_book_2010, Tauber_book_2014, 
Lindenberg_book_2019}. 
Temporal fluctuations and correlations induced by the reaction kinetics are expected to be enhanced by the presence of a 
mean-field interaction potential; this will next be quantified by evaluating the particle number variance at criticality.

\section{Particle number variance at the bicritical point: analytical approach}

Under idealized conditions, \textit{i.e.}, when the interaction parameter $w = 0$, a cumulant generating function 
representation is possible, and the inner expansion of a singular perturbation analysis allows the quantitative analysis of the 
scaling of the variance $\langle (\delta n)^2 \rangle$ for the number of particles of type X with system size (or volume) 
$\Omega$, at the bifurcation point. 
In the thermodynamic limit $(n,\Omega) \to \infty$ with fixed density $n / \Omega$, the particle number variance becomes
non-extensive at the bifurcation.
Instead, at the steady state it follows a critical power law $\langle (\delta n)^2 \rangle \sim \Omega^{\beta_0}$ with 
non-interacting scaling exponent $\beta_0 = 3 / 2$ \cite{Van_Kampen_1976, Nicolis_1977, Nicolis_1979}. 
This central result quantifies the breakdown of the central limit Gaussian approximation for large $n$ limit at criticality \cite{Mansour_1981}.

We wish to extend this result to the bicritical point of the interacting stochastic model (\ref{eq:32}). 
When $w \ne 0$, the exponentials in this master equation yield infinite polynomial series which render the generating 
function approach unfeasible. 
To progress analytically, as in the mean-field description, we introduce a minimal description that possesses the same 
qualitative behavior as the original model in the thermodynamic limit by substituting the exponentials in eq.~(\ref{eq:32})
with the common general expression $\exp\!\left( n w / \Omega \right)$ by its Taylor expansion \textit{prior to} raising
it to the power $m > 0$, 
\begin{equation}
	\exp\!\left( m n \, \frac{w}{\Omega}\, \right) = \left[ 1+ n \frac{w}{\Omega} 
	+ \mathcal{O}\left( \left(n \frac{w}{\Omega}\right)^2 \right) \right]^m ,
\label{eq:33}
\end{equation}
where the two lowest-order terms will suffice. 
This expansion could actually be terminated at any odd-order term in order to recover the bicritical behavior in the 
thermodynamic limit, with the only differences residing in shifted (non-universal) values of the ensuing critical point.  

From eqs.~(\ref{eq:32}) and (\ref{eq:33}), one may derive a minimal polynomial master equation and therefrom deduce
the scaling behavior of the particle number variance with system size $\Omega$ at the bicritical point; the difference with
respect to the non-interacting case is merely that the present model with a weak attractive mean-field potential involves 
higher-order polynomials. 
As detailed in the SM, appendix C, upon introducing a cumulant generating function and by applying a singular perturbation 
analysis with respect to a small dimensionless parameter $\epsilon = |w|  /\Omega \ll 1$, we find at the steady state in the 
thermodynamic limit $\epsilon \to 0$,
\begin{equation}
	\frac{\langle (\delta n)^2 \rangle-\langle n \rangle}{\Omega/|w|} \sim \epsilon^{- 5 / 7} \ .
\label{eq:36}
\end{equation}
Thus, at fixed $|w|$, the particle number variance diverges with system size $\Omega$ at the bicritical point as 
$\langle (\delta n)^2 \rangle \sim \Omega^\beta$ in the thermodynamic limit, with $\beta = 12 / 7$. 
In the absence of any potential ($w = 0$), or more generally, in the interacting system outside the bicritical region, in
contrast $\beta_0 = 3/2$ at the bifurcation point. 
Hence one observes a genuine change in the value of the critical scaling exponent $\beta$ due to interactions. 
Physically, this exponent difference $\Delta = 12/7 - 3/2 = 3/14$ quantifies the enhancement of fluctuation effects as the 
system is driven from the usual reactive bifurcation to the bicritical point of the stochastic model (\ref{eq:32}) that accounts 
for mean-field interactions. 
Since $\beta < 2$, we may furthermore check with the aid of eq.~(\ref{eq:32}) that the mean particle number still converges 
to its deterministic value as obtained from the solution of eq.~(\ref{eq:17}) in the thermodynamic limit, 
$\langle n \rangle = - \Omega / 2w$ as $\epsilon \to 0$ at the bicritical point in parameter space 
($\{\lambda, q\}$ = $\{0, - 1 / (4 w) \}$), as all fluctuation corrections, \textit{i.e.}, terms proportional to 
$\langle (\delta n)^2 \rangle/\langle n \rangle^2$, $\langle (\delta n)^3 \rangle/\langle n \rangle^3, \ldots$, vanish in the 
corresponding asymptotic limit \cite{Nicolis_1977}. 
Hence, in the present case, there is no fluctuation-induced shift to the mean particle number and the (mean-field) scaling 
exponent of the order parameter with respect to the control parameter is recovered. 
To check these analytical findings, we next introduce an iterative procedure to numerically solve the (full) master equation 
(\ref{eq:32}) at the steady state.

\section{Particle number variance at the bicritical point: numerical evaluation}

A natural approach to solve the master equation (\ref{eq:32}) is to employ an iterative numerical method rather than a 
standard Monte Carlo or Gillespie simulation algorithm to reach the non-equilibrium steady state which can be time-consuming 
\cite{Gillespie_1977}, since we are only interested in stationary properties here and not the transient temporal evolution. 
To that end, we set the time derivative in eq.~(\ref{eq:32}) zo zero, and obtain a general relation for the probability at the 
steady state $P(n, t \to \infty) = P_\textrm{st}(n)$ to encounter a configuration with $n$ particles of species X: 
$P_\textrm{st}(n) = P_\textrm{st}(0) \prod_{i=0}^{n-1} \left[ A (i) / B (i) \right]$, where 
\begin{equation}
\begin{split}
	A (i) &= \frac{k_{+1}}{\Omega^2} \, n_A \, i (i-1) \exp\!\left[ \frac{w}{\Omega} (2i-3) \right] + k_{-2} \, n_B \ , \\
	B (i) &= \frac{k_{-1}}{\Omega^2} \, (i+1) i (i-1) \exp\!\left[ \frac{3w}{\Omega} (i-1) \right] \\
	&\quad + k_{+2} \, (i+1) \exp\!\left( \frac{w}{\Omega} i\right) .
\label{eq:38}
\end{split}
\end{equation}
These relations can be solved iteratively. 
When $w = 0$, the particle number variance can be extracted straightforwardly, and one easily verifies that it scales with the 
volume with a scaling exponent $\beta_0 = 3/2$ at the bifurcation point and in the thermodynamic limit, as predicted. 
However, for the interacting system with $w \ne 0$, the presence of unstable steady states in the large-$n$ limit renders the 
ensuing mean and hence the variance divergent. 
Equivalently, the probability distribution $P_\textrm{st}(n)$ is inevitably non-zero for large $n$; hence, the infinite series for 
the moments extending from $n = 0$ to $\infty$ must diverge.  

To resolve this issue, we first note that the full probability distribution, defined over the entire range of values of $n$, is not in 
fact required for the evaluation of the scaling exponent $\beta$ for the particle number variance in the thermodynamic limit. 
Indeed, by taking $\Omega \to \infty$, at either critical point (bifurcation or bicritical), the stationary distributions 
$P_\textrm{st}(n)$ tend to be symmetrical about their means. 
Let us denote by $n_\textrm{max}$ the locations of the maxima of the probability distributions in the thermodynamic limit, 
and then define the \textit{reduced} variance $\langle (\delta n)^2 \rangle' = \langle n^2 \rangle' - \langle n {\rangle'}^2$ as
\begin{equation}
	\langle (\delta n)^2 \rangle' = \sum_{n=0}^{n_\textrm{max}} n^2 \, P_\textrm{st}(n) 
	- \left( \sum_{n=0}^{n_\textrm{max}} n \, P_\textrm{st}(n) \right)^{\! 2} .
\label{eq:39}
\end{equation}
At criticality, it obeys the asymptotic power law
\begin{equation}
	\lim_{\left( n, \Omega \right) \to \infty, n / \Omega \, \mathrm{finite}} \langle (\delta n)^2 \rangle' 	\sim 
	\Omega^\beta \sim \lim_{\left(n, \Omega \right) \to \infty, n / \Omega \, \mathrm{finite}} \langle (\delta n)^2 \ ,
\label{eq:40}
\end{equation}
since by symmetry, the reduced variance $\langle (\delta n)^2 \rangle' \rightarrow \langle (\delta n)^2 \rangle/2$ as both
$(n, \Omega) \to \infty$, with $n / \Omega$ held fixed and finite, and so preserves the same scaling properties with respect to 
$\Omega$ as the full particle number variance $\langle (\delta n)^2 \rangle$. 

\begin{figure}[]						
\begin{center}
\includegraphics[width=7.5cm,height=7.5cm,keepaspectratio]{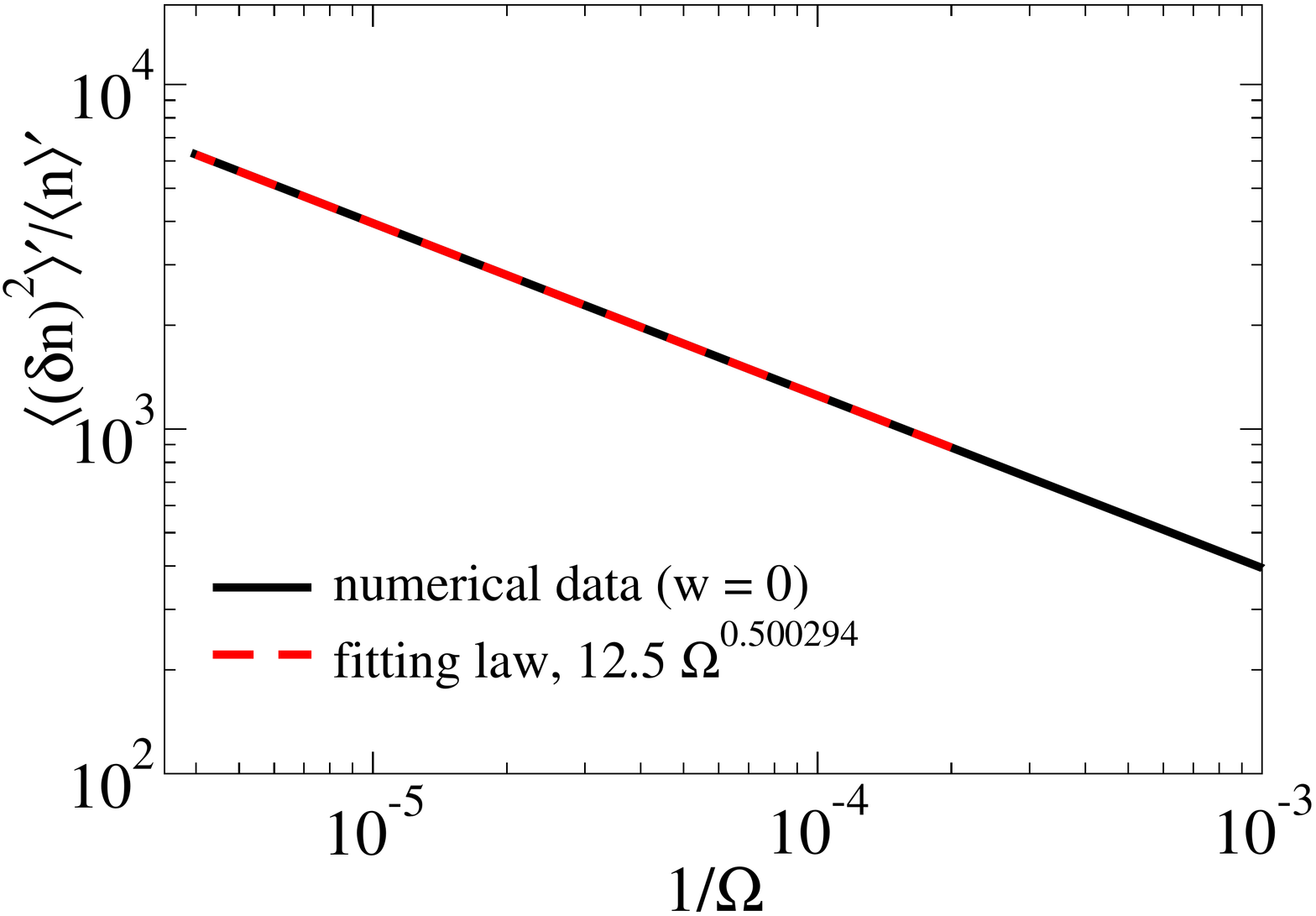}
\put(-40,29){$\displaystyle (a)$} \\
\includegraphics[width=7.5cm,height=7.5cm,keepaspectratio]{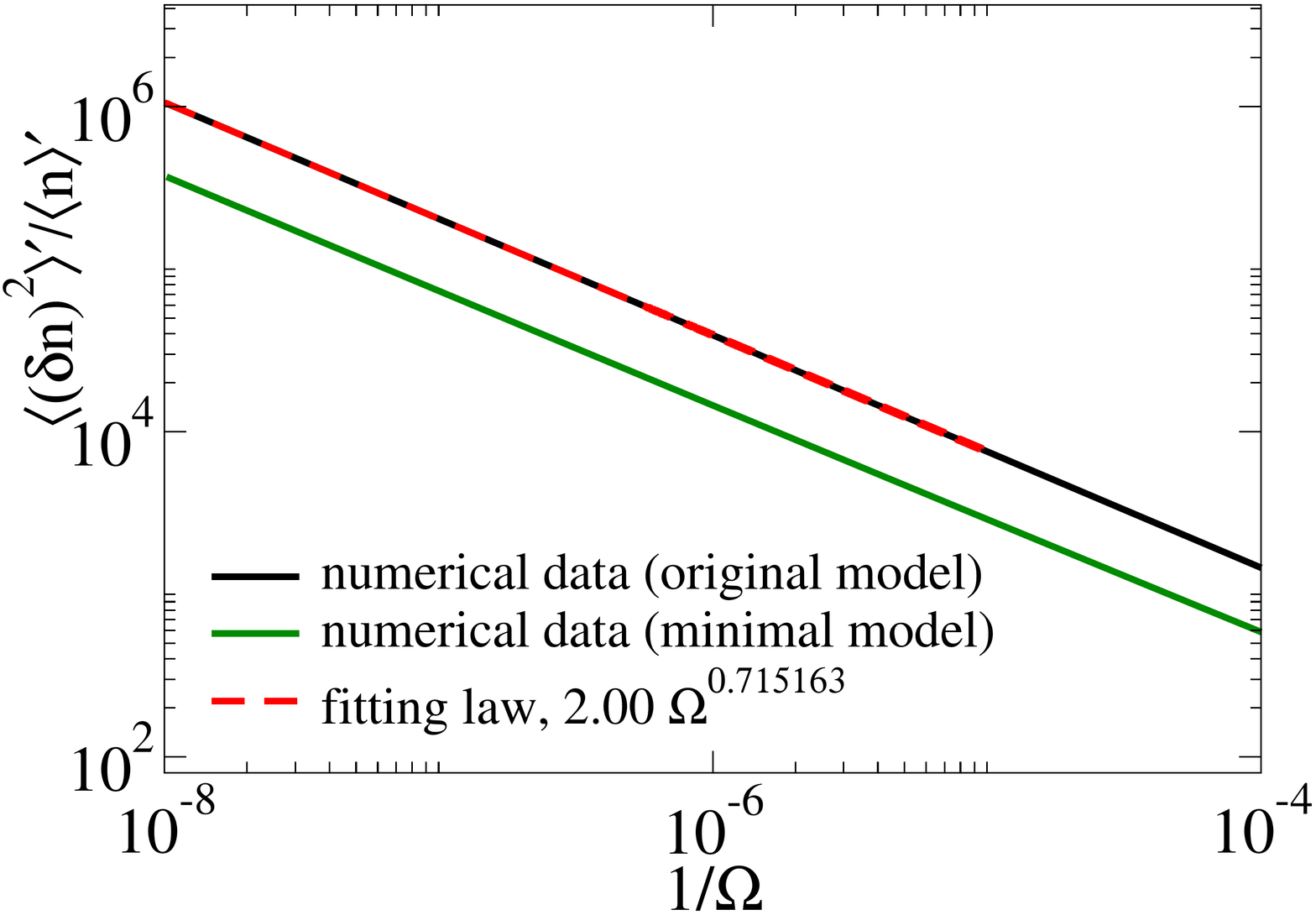}
\put(-40,29){$\displaystyle (b)$}
\caption{Rescaled reduced particle number variance $\langle (\delta n)^2 \rangle' / \langle n \rangle'$ as a function of the 
	inverse volume $1 / \Omega$, for 
	(a) $w = 0$ at the reactive bifurcation ($n_\textrm{max} = q \, \Omega$, $q=1000$); and 
	(b) $w = -0.10$ at the bicritical point, both for the original model [$n_\textrm{max} = \Omega / |w|$, $q = 1 / (e |w|)$] 
	and the minimal polynomial model [$n_\textrm{max} = \Omega / (2 |w|)$, $q = 1 / (4 |w|)$] (double-logarithmic 
	scales). 
	By fitting the numerical data that approach the thermodynamic limit, we find that the reduced variance scales with volume 	
	with scaling exponents (a) $0.500294$, and (b) $0.715163$ in excellent agreement with the theoretical predictions.} 
\label{fig:fig_4}
\end{center}
\end{figure}
The relation (\ref{eq:40}) thus allows the determination of the scaling exponent $\beta$ even when the full probability 
distribution is not accessible by iteration, such as in the present situation at the bicritical point with $w \ne 0$. 
To check the accuracy of (\ref{eq:40}), we first iteratively compute $\beta_0$ from eqs.~(\ref{eq:38}), (\ref{eq:39}) at the 
reactive bifurcation point $w = 0$, where $n_\textrm{max} = q \, \Omega$ in the thermodynamic limit, see 
fig. \ref{fig:fig_4}(a). 
For improved numerical accuracy of the fitting procedure, we plot the rescaled reduced variance 
$\langle (\delta n)^2 \rangle' / \langle n \rangle'$, where 
$\langle n \rangle' = \sum_{n=0}^{n=n_\textrm{max}} n \, P_\textrm{st}(n)$, vs. the inverse volume $1 / \Omega$.
We find this ratio to scale $\sim \Omega^{0.50...}$ in the large-$\Omega$ limit (with $n / \Omega$ fixed). 
Since $\langle n \rangle' \sim \Omega$, we deduce $\langle (\delta n)^2 \rangle' \sim \Omega^{\beta_\textrm{num}}$, 
with $\beta_\textrm{num} =1.500$, which recovers the exact scaling exponent $\beta_0 = 3/2$ derived for the (full) particle 
number variance as described by eq.~(\ref{eq:40}).

We next apply the same analysis for $w\ne 0$ at the bicritical point, where $n_\textrm{max} = \Omega / |w|$, 
fig.~\ref{fig:fig_4}(b). 
We measure $\langle (\delta n)^2 \rangle'/\langle n \rangle' \sim \Omega^{0.7151...}$ in the large-$\Omega$ limit (with 
$n / \Omega$ fixed) and thus $\langle (\delta n)^2 \rangle' \sim \Omega^{\beta_\textrm{num}}$ with 
$\beta_\textrm{num} =1.7151...$, in excellent agreement with the analytical result $\beta = 12/7 = 1.7142...$
The same effective exponent slope is found when the exponentials in the original model (\ref{eq:32}) are replaced by their 
expansions (\ref{eq:33}), as demonstrated in fig. \ref{fig:fig_4}(b). 
This confirms that the polynomial model based on the latter provides a minimal representation for the universal critical 
behavior and associated scaling properties of interest.

\section{Conclusions and prospects}

Of course, the vast majority of systems encountered in Nature deviate from ideality even in an approximate manner, since
they typically involve strongly interacting multi-component species. 
In this letter, we have introduced a simple representation of such systems to highlight the interplay between the physics of 
mutually interacting constituents (as described by an interaction Hamiltonian) and the chemistry involved when at least some 
of these species are also subject to reactions. 
Such combinations of physical and chemical interactions are characteristic of the complexity of biological systems.
Some of the ensuing features are already illuminated via the second Schl\"ogl model, a single-variable autocatalytic model 
exhibiting a bifurcation to bistability in ideal systems. 
At the mean-field level, we have shown that the presence of an attractive mean-field potential between the intermediate 
species X in this model induces an unmixing phase separation; when this transition coalesces with the reaction kinetics 
bifurcation, a bicritical point ensues. 
This bicritical point represents a unique feature of interacting particle systems subject to non-equilibrium reactions.
Hence, we devote our attention particularly to the universal properties at this bicritical point, where the effect of fluctuation are 
enhanced, as can be quantified by evaluating the particle number variance and more specifically, its scaling with the system 
size (or volume). 
To that end, we have introduced a stochastic model based on the chemical master equation that is built on a generalized 
detailed balance condition and incorporates transition rates consistent with the transition state theory of reaction-limited 
processes; the weak attractive particle interactions are treated in a mean-field setting.
We have constructed a minimal polynomial model and demonstrated that it captures the essential universal critical behavior of 
the interacting system.
Our analysis yields that the particle number variance scales with the system size with a scaling exponent $\beta = 12 / 7$ in 
the thermodynamic limit, distinct from $\beta_0 = 3 / 2$ found at the reactive bifurcation point. 
The deviation $\Delta = 12/7 - 3/2 = 3/14$ precisely quantifies the enhancement of (internal) noise induced by the combined 
effect of the reaction kinetics and of the mean-field interactions at criticality. 
This analytical result is in excellent agreement with our numerical estimate, $\beta_{num} = 1.7151...$, deduced from 
iteratively solving the master equation (\ref{eq:32}). 

Both mean-field and stochastic models considered here are zero-dimensional, devoid of any spatial structure. 
Hence, a natural extension of this work would be to provide a complete description of the critical behavior with diffusive 
transport and taking into account the dependence of the interaction potential on the relative positions between lattice sites on 
a $d$-dimensional lattice \cite{Glotzer_1994, Glotzer_1995, Hildebrand_1996, Decker_2006}. 
One possible way to fix the transition rates in such a more general framework is to couple Kawasaki exchange dynamics to 
non-conserved Glauber dynamics to respectively model diffusive transport and chemical reactions in a phase-separating 
environment \cite{Glotzer_1994, Glotzer_1995}. 
One may anticipate modified scaling laws and novel dynamical scenarios related to non-equilibrium pattern formation, which 
could drastically deviate from the mean-field approach, especially close or at criticality and/or in low-dimensional media 
\cite{Grassberger_1982, Cross_1993, Hohenberg_1997, Odor_2004, Tauber_book_2014, Tauber_2017}. 
Furthermore, an extension of this work to two-and three-variable autocatalytic systems to revisit symmetry-breaking 
instabilities to spatio-temporal chaos and chemical oscillations could also be of great interest in the future 
\cite{Gaspard_book_2022}. 
We hope that this present work will provide a first theoretical basis for many more theoretical and (hopefully) experimental 
investigatons devoted to quantifying fluctuation effects in multi-component interacting systems that are actively influenced by 
the presence of chemical reactions.

\acknowledgments
R.T. gratefully acknowledges financial support from the BAEF / Francqui foundations.

\section{Supplementary material}

\subsection{Appendix A: Derivation of the mean-field model}

The governing equation for the number density $c$ of particles of species X, under the assumption that the 
temperature $T$ is uniform (fixed by a heat bath) and in the absence of any bulk flow, reads
\begin{equation}
	\frac{\partial c}{\partial t} = - \nabla \, \underline{J}^{\mathrm{dif}}  +  \sigma,
\label{eq:3}
\end{equation}
where $\underline{J}^{\mathrm{dif}}$ and $\sigma$ are the (vectorial) diffusive flux and (scalar) production 
source of species X, respectively. 
Equation~(\ref{eq:3}) can be rigorously obtained from the conservation of mass of the particles of type X flowing
through the surface of a system element of volume $\Omega$ by means of Gauss’ divergence theorem. 
Closing the mathematical description requires constitutive relations for the fields $\underline{J}^{\mathrm{dif}}$ 
and $\sigma$. 
The former can be derived in the theoretical framework of non-equilibrium thermodynamics \cite{Groot_book_1984}. 
Within the linear response regime, \textit{i.e.}, close to thermal equilibrium, the particle current 
$\underline{J}_{k}^{\mathrm{dif}}$ of any component $k$ at uniform temperature is related to the 
thermodynamic force associated with chemical potential gradients $\nabla {\mu_j}$ by the phenomenological 
Onsager coefficients $L_{jk}$ via \cite{Li_1981_a, Groot_book_1984}
\begin{equation}
	\underline{J}_{k}^{\mathrm{dif}} =  -\sum^{\mathcal{C}}_{j} \frac{L_{jk}}{T} \, \nabla {\mu_j} \ ,
\label{eq:4}
\end{equation}
\noindent
where $\mathcal{C}$ represents the total number of species involved. 
Neglecting cross-diffusion terms, the diffusive flux $\underline{J}^{\mathrm{dif}} $ of species X is then related to 
its chemical potential gradient $\nabla {\mu_X}$ through the diagonal Onsager coefficient $L_{XX} = L$,
\begin{equation}
	\underline{J}^{\mathrm{dif}} = - \frac{L}{T} \, \nabla {\mu_X} \ .
\label{eq:5}
\end{equation}

We next proceed to provide a model for the kinetic source term $\sigma$, which we define as the total production
rate of component X in a system of $\alpha = 1, \ldots, M$ (reversible) chemical reactions: 
$\sigma  = \sum_{\alpha=1}^{M} \nu_{\alpha} \sigma_{\alpha}$, where the net stochiometric coefficient 
$\nu_{\alpha} = \nu_{\alpha}^{(p)} - \nu_{\alpha}^{(r)}$ is given by the difference between the stochiometric 
coefficients of component $\alpha$ on the product $(p)$ and reactant $(r)$ sides of the reaction, and the net 
reaction rate $\sigma_{\alpha} = \sigma_{+ \alpha} - \sigma_{- \alpha}$ similarly represents the difference 
between the rates of the forward $(+)$ and backward $(-)$ reactions. 
We assume the system to be connected to particle reservoirs (chemostats) for reactant A and product B. 
A generalized detailed balance condition then holds for any species $\alpha$, provided 
\cite{Weber_2019, Zwicker_2022},
\begin{equation}
	\frac{\sigma_{+ \alpha}}{\sigma_{- \alpha}} = \mathrm{exp}\!
	\left( - \, \frac{\nu_{\alpha \textrm{X}} \, \mu_{X} + \Delta {\mu}_{\alpha}}{k_\mathrm{B} T} \right) ,
\label{eq:6}
\end{equation}
where $k_\mathrm{B}$ is Boltzmann’s constant and $\Delta {\mu}_{\alpha} = 
\nu_{\alpha \textrm{B}} \, {\mu}_\textrm{B} - \nu_{\alpha_\textrm{A}} \, {\mu}_\textrm{A}$ denotes the net
chemical work done by the reservoirs on the system, with ${\mu}_\textrm{B}$ and ${\mu}_\textrm{A}$ 
representing the chemical potentials of the chemostatted product B and reactant A, respectively. 
In the absence of any reservoir, \text{i.e.}, when $\nu_{\alpha B} = 0 = \nu_{\alpha A}$, the system in contact 
with the heat bath relaxes toward thermal equilibrium described by the canonical ensemble, and we recover from 
eq.~(\ref{eq:6}) the standard equilibrium detailed rate balance. 
The presence of external particle reservoirs with fixed $\Delta {\mu}_{\alpha} \ne 0$ induces a net particle flow 
through the system and prevents relaxation to equilibrium by violating the detailed-balance condition, whence the 
system is driven toward possibly a non-equilibrium steady state (if it exists).

To fix the individual forward and backward rates $\{\sigma_{+\alpha}, \sigma_{-\alpha}\}$ in eq.~(\ref{eq:6}), 
we invoke the kinetic theory of transition states for reaction-limited processes \cite{Hanggi_1990}. 
Within this framework, the reaction rates are determined by the typical time scales for an activated complex 
formed by the reactant particles to cross the free-energy barrier along the reaction pathway; hence the rates 
generally depend on composition. 
Yet we posit the simplifying assumption that such a composition dependence does not apply here; then 
according to transition state theory in accordance with the generalized thermodynamic constraint (\ref{eq:6}),
\begin{equation}
\begin{split}
	&\sigma_{+ \alpha} = g_{\alpha} \, \mathrm{exp}\!\left[ \left( \nu_{\alpha \textrm{X}}^{(r)} \mu_\textrm{X} 
	+ \nu_{\alpha \textrm{A}} \, {\mu}_\textrm{A} \right) / k_\mathrm{B} T \right] , \\
	&\sigma_{- \alpha} = g_{\alpha} \, \mathrm{exp}\!\left[ \left( \nu_{\alpha \textrm{X}}^{(p)} \mu_\textrm{X} 
	+ \nu_{\alpha \textrm{B}} \, {\mu}_\textrm{B} \right) / k_\mathrm{B} T \right] , 
\label{eq:7}
\end{split}
\end{equation}
with constant prefactors $g_{\alpha} > 0$ that do not dependent on the densities. 

The set of chemical potentials $\{\mu_\textrm{X}, {\mu}_\textrm{A}, {\mu}_\textrm{B} \}$ can be computed from 
the (classical) canonical partition function $Z(T, \Omega, \{ n_j \})$ for the system with size $\Omega$ at temperature 
$T$ via \cite{Schwabl_book_2006, Kardar_book_2007}
\begin{equation}
	\mu_j = - k_\mathrm{B} T 
	\left( \frac{\partial \, \mathrm{ln} \, Z}{\partial n_j} \right)_{T, \Omega, n_{k \ne j}} ,
\label{eq:9}
\end{equation}
with $n_j$ indicating the particle number of species $j$ ($=$ X, A, B here). 
Pairwise interactions between point-like particles of either species give in the mean-field limit 
\begin{equation}
	Z = \frac{1}{\prod_{j}^{\mathcal{C}} n_j!} \ \prod_{j}^{\mathcal{C}} \frac{\Omega}{\lambda_j(T)^d} \  
	\mathrm{exp}\!\left(- \frac{U_T}{k_\mathrm{B} T} \right) 
\label{eq:10}
\end{equation}
in $d$ dimensions, with the thermal de~Broglie wavelength 
$\lambda_j(T) = h / (2 \pi \, m_j k_\mathrm{B} T)^{1/2}$ for non-relativistic particles of mass $m_j$ ($h$ is 
Planck's constant), and the total interaction potential energy
\begin{equation}
	U_{T} =\frac{1}{2} \sum_{j,k}^{\mathcal{C}} n_j n_k \, u_{jk} \ ,
\label{eq:11}
\end{equation}
where $u_{jk}$ denotes the attractive component of the interaction potential between species $j$ and $k$ averaged 
over the volume $\Omega$,
\begin{equation}
	u_{jk} = \frac{1}{\Omega} \int_{\Omega} d\underline{r}_{jk} \, U_{jk} \left(\underline{r}_{jk}\right) = 			
	\frac{\epsilon_{jk}}{\Omega} < 0 \ .
\label{eq:12}
\end{equation}
Here, $\underline{r}_{jk}$ is the relative position vector between two particles of species $j$ and $k$; and we have 
introduced $\epsilon_{jk}$ as the corresponding volume-integrated mean-field potential. 
Equations~(\ref{eq:9})--(\ref{eq:12}) yield the chemical potential of species $j$ 
\begin{equation}
	\mu_j =  \mu_j^{0} (T) + k_\mathrm{B} T \, \ln c_j + \sum_{k}^{\mathcal{C}} \epsilon_{jk} c_k \ , 
\label{eq:13} 
\end{equation}
where $\mu_j^{0} (T) = k_\mathrm{B} T \ln \lambda_j(T)^{d}$ only depends on temperature and 
$c_k = n_k / \Omega$.

To simplify further, we assume only the mutual interactions between particles of species X to be relevant, and neglect 
any other pair interactions, $\epsilon_{jk} = 0$ for $j \ne k$ and 
$\epsilon_\textrm{AA} = 0 = \epsilon_\textrm{BB}$. 
This turns out sufficient to model the (mean-field) critical behavior of interest. 
We thus obtain
\begin{equation}
\begin{split}
	&\mu_\textrm{A,B} = \mu_\textrm{A,B}^{0} (T) + k_\mathrm{B} T \, \ln c_\textrm{A,B} \ , \\
	&\mu_{X} = \mu_{X}^{0} (T) + k_\mathrm{B} T \left( \ln c + w \, c \right) ,
\label{eq:15} 
\end{split}
\end{equation}
where $w =\epsilon_\textrm{XX} / k_\mathrm{B} T < 0$ is the rescaled volume-integrated mean-field potential.
As explained in the main text, the attractive nature of the mean-field interactions $w < 0$ is a necessary condition 
for the emergence of an unmixing phase separation transition and hence for the presence of a bicritical point. 
With eqs.~(\ref{eq:5}), (\ref{eq:7}), (\ref{eq:15}) and $\sigma = \sigma_{+1} - \sigma_{-1} - \sigma_{+2} 
+ \sigma_{-2}$, the evolution equation (\ref{eq:3}) for the number density of species X finally becomes
\begin{equation}
\begin{split}
	\frac{\partial c}{\partial t} &= D \nabla \left[ (1+wc) \nabla c \right] + k_{+1} \, c_\textrm{A} \, [c \, \gamma(c)]^2 
	\\ &\quad - k_{-1} \, [c \, \gamma(c)]^3 - k_{+2} \, c \, \gamma(c) + k_{-2} \, c_\textrm{B} \ ,
\label{eq:16}
\end{split}
\end{equation}
eq.~(2) of the main text.
Here $\gamma (c) = \mathrm{exp}(wc)$ denotes the density-dependent functional correction due to the attractive 
mean-field interactions, $D = L \, k_\mathrm{B} / c$ is the diffusion coefficient, assumed to be constant here, and the 
kinetic rate parameters are given by 
\begin{equation}
\begin{split}
	&k_{+\alpha} = g_{\alpha} \, \mathrm{exp}\!\left[ \left( \nu_{\alpha \textrm{X}}^{(r)} \, \mu_\textrm{X}^{0} 
	+ \nu_{\alpha \textrm{A}} \, \mu_\textrm{A}^{0} \right) / k_\mathrm{B} T \right] , \\
	&k_{-\alpha} = g_{\alpha} \, \mathrm{exp}\!\left[ \left(\nu_{\alpha \textrm{X}}^{(p)} \, \mu_\textrm{X}^{0} 
	+ \nu_{\alpha \textrm{B}} \, \mu_\textrm{B}^{0} \right) / k_\mathrm{B} T \right] . 
\end{split}
\end{equation}

\subsection{Appendix B: Derivation of the stochastic model}

We describe an elementary reactive process as a death-birth Markov process. 
For the second Schl\"ogl model, the evaluation equation for the probability $P(n, t)$ to find a configuration with $n$ 
particles of species X at time $t$ in a non-spatial (zero-dimensional) setting takes the form
\begin{equation}
\begin{split}
	\frac{\partial P(n,t)}{\partial t} &= \big[ W_{+1}(n \vert n-1) + W_{-2}(n \vert n-1) \big] P(n-1,t) \\
	&\ + \big[ W_{-1}(n \vert n+1) + W_{+2}(n \vert n+1) \big] P(n+1,t) \\
	&\ - \big[ W_{+1}(n+1 \vert n) + W_{-1}(n-1 \vert n) \\
	&\qquad + W_{+2}(n-1 \vert n) + W_{-2}(n+1 \vert n) \big] P(n,t) \ , 
\label{eq:27}
\end{split}
\end{equation}
where $W_{+\alpha}$ and $W_{-\alpha}$ represent the transition rates of the forward and backward processes
for the reversible chemical reaction labeled by $\alpha$, respectively, with the convention that $(\nu \vert \nu')$ 
illustrates the transition from a configuration with $\nu'$ to $\nu$ particles, where $\nu$, $\nu' = n-1$, $n$, or 
$n+1$.

In order to determine the transition rates and hence close eq.~(\ref{eq:27}), we impose a generalized detailed 
balance condition akin to eq.~(\ref{eq:6}) on the ratio 
\begin{equation}
\begin{split}
	& \frac{W_{+\alpha} (\ldots \vert n)}{W_{- \alpha} (\ldots \vert n+\Delta \nu_{\alpha \textrm{X}})} = \\ 
	&\quad = \mathrm{exp}\!\left(- \frac{F(n+\Delta \nu_{\alpha \textrm{X}}) - F(n) + \Delta \mu_{\alpha}}
	{k_{\mathrm{B}}T} \right) ,
\label{eq:28}
\end{split}
\end{equation}
where $\Delta \nu_{\alpha \textrm{X}} = \nu_{\alpha \textrm{X}}^{(p)} - \nu_{\alpha \textrm{X}}^{(r)}$, 
whereas the individual rates are fixed by the kinetic transition state theory, see eq.~(\ref{eq:7}),
\begin{equation}
\begin{split}
	&W_{+\alpha} (\ldots \vert n) = g_{\alpha} \, \mathrm{exp}\!\left[ \frac{F(n) 
	- F\big( n- \nu_{\alpha \textrm{X}}^{(r)} \bigr) + \nu_{\alpha \textrm{A}} \, \mu_\textrm{A}}
	{k_{\mathrm{B}} T} \right] , \\
	&W_{- \alpha}(\ldots \vert n+\Delta \nu_{\alpha \textrm{X}}) = \\
	&\ g_{\alpha} \, \mathrm{exp}\!\left[ \frac{F(n+\Delta \nu_{\alpha \textrm{X}}) 
	- F\bigl( n+\Delta \nu_{\alpha \textrm{X}} - \nu_{\alpha \textrm{X}}^{(p)} \bigr) 
	+ \nu_{\alpha \textrm{B}} \, \mu_\textrm{B}}{k_{\mathrm{B}} T} \right] , 
\label{eq:30}
\end{split}
\end{equation}
with a prefactor $g_{\alpha}$ that we take to be state-independent. 

The two conditions (\ref{eq:28}) and (\ref{eq:30}), the stochastic analogs of eqs.~(\ref{eq:6}) and (\ref{eq:7}), 
ensure that as the system relaxes to thermal equilibrium, the canonical Boltzmann--Gibbs distribution is recovered 
in the long-time limit, when there is no external drive ($\Delta \mu_{\alpha} = 0$ $\forall \alpha$).  
Moreover, the temporal evolution of the solute X density $c$ is consistent with transition state theory 
\cite{Schmittmann_1990, Julicher_1997, Seifert_2005, Pietzonka_2014, Miangolarra_2023}. 
We note that similar expressions for the transition rates have been considered in various instances, \textit{e.g.}, 
for the study of molecular motors \cite{Julicher_1997, Seifert_2005, Pietzonka_2014} and more generally, for 
driven diffusive systems \cite{Schmittmann_1990, Schmittmann_book_1995, Marro_book_1999}. 

The discrete free energy $F (\nu) = - k_{\mathrm{B}} T \ln Z(\nu)$ follows from the canonical partition function 
$Z(\nu)$ given by eq.~(\ref{eq:10}) for an interacting gas system.
Upon averaging the weakly attractive interaction contribution in the Hamiltonian over the volume $\Omega$ as in 
(\ref{eq:10}), the total mean-field interaction energy takes the general form (for arbitrary integers $n_j$)
\begin{equation}
	U_{T} = \frac12 \sum_{j \ne k}^{\mathcal{C}} n_j n_k \, u_{jk} 
	+ \frac12 \sum_{j = k}^{\mathcal{C}} n_j (n_k - 1) \, u_{jk} \ .
\label{eq:31}
\end{equation}
Combining eqs.~(\ref{eq:27})--(\ref{eq:31}) yields eq.~(9) of the main text,
\begin{equation}
\begin{split}
	&\frac{\partial P(n,t)}{\partial t} = 
		k_{+2} \, (n+1) \, \mathrm{exp}\!\left( \frac{w}{\Omega} n \right) P(n+1,t) \\
	&\ + \frac{k_{+1}}{\Omega^2} \, n_\textrm{A} (n-1) (n-2) \, \mathrm{exp}\!\left[ \frac{w}{\Omega} (2n-5) \right] 
		P(n-1,t) \\
	&\ + \frac{k_{-1}}{\Omega^2} \, (n+1) n (n-1) \, \mathrm{exp}\!\left[ \frac{3w}{\Omega} (n-1) \right] P(n+1,t) \\
	&\ + k_{-2} n_B \, P(n-1,t) - k_{-2} \, n_\textrm{B} \, P(n,t)  \\
	&\ - \frac{k_{+1}}{\Omega^2} \, n_\textrm{A} n(n-1) \, \mathrm{exp}\!\left[ \frac{w}{\Omega} (2n-3) \right] P(n,t) \\
	&\ - \frac{k_{-1}}{\Omega^2} \, n (n-1) (n-2) \, \mathrm{exp}\!\left[ \frac{3w}{\Omega} (n-2) \right] P(n,t) \\
	&\ - k_{+2} \, n \, \mathrm{exp}\left[ \frac{w}{\Omega} (n-1) \right] P(n,t) \ .
\label{eq:32}
\end{split}
\end{equation}

\subsection{Appendix C: Variance at the bicritical point in the thermodynamic limit}

Let us define the generating function $F(s) = \sum_{n=0}^{\infty} s^n P(n)$. 
From eqs.~(13) and (14) of the main text and at the steady state, a (non-linear) ordinary differential equation (ODE) of sixth 
order with coefficients of degree 5 or less in the variable $s$ can be derived for $F(s)$, which takes the form 
$\left( d^6 F(s) / ds^6 \right) w^4 \, \Omega^{-4} + \ldots + \ldots \Omega^2 \, w^{-2} F(s) = 0$. 
Upon introducing a cumulant generating function representation, $\upsilon (s) = d\varphi(s) / ds$, 
$F(s) = \exp\! \left( \Omega \varphi (s) / |w| \right)$ and defining $\epsilon = |w| / \Omega$, the corresponding ODE for 
$F(s)$ is replaced by a fifth-order ODE for $\upsilon (s)$ with coefficients of degree 5 or less in terms of the dimensionless perturbation parameter $\epsilon$. 
It is of the form $\left(d^5 \upsilon / ds^5 \right) \epsilon^5 + \ldots \upsilon^6 + \ldots + \ldots s^{-2} = 0$, where the
omitted terms are not required for the present demonstration to recover eq.~(11) of the main text. 
The thermodynamic limit corresponds to $\epsilon \to 0$, which is equivalent to $\Omega \to \infty$ and/or $|w| \to 0$.

The ODE for $\upsilon (s)$ serves as the starting point for a singular perturbation analysis from which the inner expansion 
gives access to the effect of fluctuations at the interacting bicritical point. 
To this end, as described in ref.~\cite{Nicolis_1977}, we perform a scale transformation on the ODE for $\upsilon (s)$, 
namely $(i)$ we stretch the vicinity of $s=1$, $s = 1 + \epsilon^{a} \eta$ ($a >0$), and $(ii)$ we expand the solution
$\Upsilon(\eta) = \upsilon (s)$ in fractional powers: 
$\Upsilon  = 1 + \epsilon^{c} \, \Upsilon_{1} + \epsilon^{2c} \, \Upsilon_{2} + \ldots$, with $c >0$. 
We thus obtain an ODE for $\Upsilon(\eta)$ which conserves the order 5 and where the coefficients in $\epsilon$ are of a 
degree that is a function of $a$ and $c$, $\left( d^5 \Upsilon / d\eta^5 \right) \epsilon^{5-5a} + \ldots \Upsilon^6 + \ldots
+ (\dots \epsilon^{a} \eta + \ldots \epsilon^{2a} \eta^2 + \ldots) = 0$. 
Truncating at $\Upsilon = 1 + \epsilon^{c} \, \Upsilon_{1} + \mathcal{O}(\epsilon^{2c})$, one arrives at a first scaling relation by balancing the highest-derivative term of order $\epsilon^{5-5a+c}$ with the independent term of order 
$\epsilon^{a}$, which gives $5 (1-a) + c = a$. 
A second scaling relation follows from by balancing one of the two latter terms with the sixth-order polynomial term (one 
may check that terms of order $\epsilon^{n c}$ with integer $n < 6$ are sub-leading in the thermodynamic limit). 
This procedure finally yields $a = 5 (1-a) + c = 6c$, which results in $a =  6 / 7$ and $c =  1 / 7$. 

The particle number variance can be evaluated by noting that
\begin{equation}
	\frac{\langle (\delta n)^2 \rangle-\langle n \rangle}{\Omega/|w|} = \left(\frac{d\Upsilon}{ds} \right)_{s=1} .
\label{eq:35}
\end{equation}
To order $\epsilon^{c}$, and if we assume that $\Upsilon_1 (\eta) \sim \eta^{b}$ in the vicinity of $\eta = 0$, we find 
that $(d\Upsilon/ds)_{s=1} \sim b \, \epsilon^{c-ab} (s-1)^{b-1}$. 
To extract $b$, we see that as $s \to 1$ for $b<1$, we obtain a divergent probability which violates its normalizability
condition, while $b >1$ would imply identical mean and variance which is generically not true in nonlinear systems and/or at 
criticality; hence the only admissible solution corresponds to $b = 1$. 
Consequently with $c - a = - 5 / 7$,
\begin{equation}
	\frac{\langle (\delta n)^2 \rangle - \langle n \rangle}{\Omega/|w|} \sim \epsilon^{-5/7} \ ,
\label{eq:36}
\end{equation}
which gives eq.~(11) of the main text.

\end{document}